\documentclass[12pt]{article}
\usepackage{amsmath}
\usepackage{graphicx}
\usepackage{subfigure}
\usepackage{hyperref}

\begin{document}

\author{Ernst Trojan and George V. Vlasov \and \textit{Moscow Institute of Physics
and Technology} \and \textit{PO Box 3, Moscow, 125080, Russia}}
\title{Thermodynamics of absolute stiff matter}
\maketitle

\begin{abstract}
The pressure, particle number density and heat capacity of 'absolute stiff'
matter ($P=E$) at finite temperature are obtained for fermions and bosons.
The behavior of 'absolute stiff' medium depends on the characteristic
temperature $T_c=6\pi ^2n/\left( \gamma m^2\right) $, and low temperature
expansion ($T\ll T_c$) is considered. We also find that the equation of
state $P=wE$ at arbitrary constant $w$ can be modeled by an ideal gas of
quasi-particles with the energy spectrum $\varepsilon _p=ap^{wq}$ in $q$%
-dimensional space.
\end{abstract}

\section{Introduction}

The 'absolute stiff' matter ($P=E$) can be a Fermi or Bose gas of particles
with the energy spectrum $\varepsilon_p\sim p^{q}$ in $q$-dimensional space.
We obtain its pressure, particle number density and heat capacity at finite
temperature. The behavior of 'absolute stiff' medium is determined by
characteristic temperature $T_c\sim n$. At low temperature the heat capacity
obeys the linear law $C_V\sim T$ for both fermionic and bosonic matter, the
pressure of 'absolute stiff' fermions $P\sim n^2+O(T^2)$, while the
'absolute stiff' Bose gas never reveals Bose-Einstein condensation.

The equation of state (EOS) is a fundamental characteristic of material. It
is a functional link $P\left( E\right) $\ between the pressure $P$\ and the
energy density $E$. Its knowledge allows to predict the behavior of matter
and calculate parameters of astrophysical object that contain this mater.
The EOS can be given in the form 
\begin{equation}
P=wE  \label{xi}
\end{equation}
where $w$ is dimensionless parameter that, in general, depends on $E$.
Particularly, the EOS of ideal gas of non-relativistic particles has
constant $w=2/3$. The EOS with $w=1/3$ describes radiation, the EOS of dust
has $w=0$, while tachyon matter admits $P>E$ \cite{TV2011c,TV2011d}\textrm{. 
}One of the most exotic examples 
\begin{equation}
P=E  \label{sti}
\end{equation}
corresponds to the so-called 'absolute stiff' matter, that may appear in
various problems of astrophysics. Particularly, it is often considered in
the interiors of neutron stars \cite{RR74,KG96,O} and problems of cosmology 
\cite{BE,BFM}. However, it is still uncertain what physical particles can
form this material. It is clear that free particles with the energy spectrum 
\begin{equation}
\varepsilon _p=\sqrt{p^2+m^2}  \label{pro}
\end{equation}
cannot constitute 'absolute stiff' matter, and only interaction between
nucleons results in the EOS very close to (\ref{sti}).

The estimation of interaction is the main difficulty because it requires
solution of the quantum many-body problem. It is often performed in the
frames or Hartree-Fock approximation. An alternative method is formulated in
the density functional theory \cite{DG90}, where a system of interacting
particles is modeled by a system of free hypothetical quasi-particles moving
in some external field. Although the system of free non-interacting
quasi-particles is considered as an ideal gas, their energy spectrum differs
from the energy spectrum of free particles (\ref{pro}). The role of
interaction can be reflected in the effective mass and effective chemical
potential as it is done in the Walecka nuclear model \cite{W}, but in
general the energy spectrum of quasi-particles may differ sufficiently from (%
\ref{pro}). Of course, it is no more than a model because such hypothetical
quasi-particles do not exist in nature, but this model is helpful for
calculations of the EOS.

In the present paper we use this model of free quasi-particles for
description of exotic forms of matter that appear in astrophysical problems.
We consider an ideal gas with EOS $P=wE$ (\ref{xi}) at constant $w$ and
establish the energy spectrum of quasi-particles that can constitute this
matter. We also study thermodynamical properties of Fermi and Bose gases
with 'absolute stiff' EOS (\ref{sti}) and outline constraints applicable to
neutron stars.

The standard relativistic units $c_{light}=\hbar =k_B=1$\ are used in the
paper.

\section{Thermodynamical functions}

Consider an ideal gas of free particles with the single-particle energy
spectrum $\varepsilon _p$ at finite temperature $T$ and in a $q$-dimensional
space. Let $\mu $ be the chemical potential of this system. The pressure $P$%
, energy density $E$, entropy density $S$, and the particle number density $%
n $ are determined by the standard formulas \cite{K89b} 
\begin{equation}
P=-T\ln Z  \label{p}
\end{equation}
\begin{equation}
E=\frac \gamma {\left( 2\pi \right) ^q}\int f_p\varepsilon _pd^qp  \label{en}
\end{equation}
\begin{equation}
n=\frac \gamma {\left( 2\pi \right) ^q}\int f_p\,d^qp  \label{n}
\end{equation}
where 
\begin{equation}
\ln Z=\mp \frac \gamma {\left( 2\pi \right) ^q}\int \ln \left\{ 1\pm \exp
\left[ (\varepsilon _p-\mu )/T\right] \right\} d^qp  \label{z}
\end{equation}
is the statistical sum, 
\begin{equation}
f_p=\frac 1{\exp \left[ (\varepsilon _p-\mu )/T\right] \pm 1}  \label{f}
\end{equation}
is the distribution function, and the sign ''$+$''or ''$-$'' corresponds to
fermions and bosons. The volume of $q$-dimensional hypersphere is defined as 
\begin{equation}
d^qp=\frac{q\pi ^{q/2}}{\Gamma \left( \frac q2+1\right) }p^{q-1}dp
\label{imp}
\end{equation}
Partial integration of (\ref{z}) and its substitution in (\ref{p}) yields 
\begin{equation}
P=\frac \gamma {\left( 2\pi \right) ^q}\frac{\pi ^{q/2}}{\Gamma \left( \frac
q2+1\right) }\int f_p\frac{\partial \varepsilon _p}{\partial p}p^qdp
\label{p0}
\end{equation}
For example, in $3$ dimensions 
\begin{equation}
d^3p=4\pi p^2dp  \label{imp3}
\end{equation}
and 
\begin{equation}
P=\frac \gamma {6\pi ^2}\int f_p\frac{\partial \varepsilon _p}{\partial p}%
p^3dp  \label{p03}
\end{equation}

Let us imagine that the medium with EOS $P=wE$ is an ideal gas of free
quasi-particles with the energy spectrum $\varepsilon _p$. From (\ref{xi}), (%
\ref{en}) and (\ref{p0}) we get equation: 
\begin{equation}
P-\xi E=\frac \gamma {\left( 2\pi \right) ^q}\frac{q\pi ^{q/2}}{\Gamma
\left( \frac q2+1\right) }\int f_p\left( \frac pq\frac{d\varepsilon _p}{dp}%
-w\varepsilon _p\right) p^{q-1}dp=0  \label{zer}
\end{equation}
whose solution is 
\begin{equation}
\varepsilon _p=ap^{wq}  \label{e0}
\end{equation}
where $a$ is an arbitrary constant.

It differs from the standard single-particle energy spectrum of free
particles (\ref{pro}) and the objects with the energy spectrum (\ref{e0})
should be referred as quasi-particles or excitations. Particularly, the
energy spectrum in the form 
\begin{equation}
\varepsilon _p=ap^q  \label{e1}
\end{equation}
belongs to the 'absolute stiff' matter (\ref{sti}), while the exotic matter
with $P=-E$ is composed of quasi-particles with the energy spectrum 
\begin{equation}
\varepsilon _p=\frac a{p^q}  \label{e-1}
\end{equation}
The dust matter with $P=0$ corresponds to $\varepsilon _p=a=\mathrm{const}$,
and massless particles with the energy spectrum 
\begin{equation}
\varepsilon _p=cp  \label{eph}
\end{equation}
always have the same EOS 
\begin{equation}
P=\frac Eq  \label{eos3}
\end{equation}

The energy spectrum of 'absolute stiff' matter (\ref{e1}) in 3-dimensional
space has the form 
\begin{equation}
\varepsilon _p=\frac{p^3}{m^2}  \label{e11}
\end{equation}
while in 2D space it is 
\begin{equation}
\varepsilon _p=\frac{p^2}{2m}  \label{e112}
\end{equation}
and in 1D space it is 
\begin{equation}
\varepsilon _p=cp  \label{e111}
\end{equation}
where constant $m$ is the mass parameter. Formula (\ref{e112}) implies that
the ordinary particles in a thin film do form an 'absolute stiff' matter (%
\ref{sti}), the fact already known in statistical mechanics of
non-relativistic gases in two dimensions \cite{A92,T94}. Phonon-like
excitations (\ref{e111}) in a thin channel form an ''absolute stiff' matter.

Substituting the energy spectrum of 'absolute stiff' material (\ref{e1}) in (%
\ref{n}) and (\ref{en}), we have 
\begin{equation}
n=\Sigma _q\frac Ta\int\limits_0^\infty \frac{dx}{\exp \left( x-\mu
/T\right) +\sigma }  \label{ni}
\end{equation}
\begin{equation}
E=P=\Sigma _q\frac{T^2}a\int\limits_0^\infty \frac{xdx}{\exp \left( x-\mu
/T\right) +\sigma }  \label{pi}
\end{equation}
where 
\begin{equation}
\Sigma \left( q\right) =\frac \gamma {\left( 2\pi \right) ^q}\frac{\pi ^{q/2}%
}{\Gamma \left( \frac q2+1\right) }  \label{sigm}
\end{equation}
\begin{equation}
x=\frac{\varepsilon _p}T  \label{xxx}
\end{equation}
and $\sigma =\pm 1$ (upper sign for fermions, lower sign for bosons). Hence,
we get universal formulas for the particle number density 
\begin{equation}
n=\Sigma _q\frac T{a\sigma }\ln \left[ 1+\sigma \exp \left( \frac \mu
T\right) \right]  \label{na}
\end{equation}
and the pressure 
\begin{equation}
E=P=\Sigma _q\frac 1{a\sigma }\left\{ \frac{\mu ^2}2+\left( \frac{\ln
^2\sigma }2+\frac{\pi ^2}6\right) T^2+T^2\mathrm{dilog}\left[ 1+\frac
1\sigma \exp \left( \frac{-\mu }T\right) \right] +\mu T\ln \sigma \right\}
\label{pa}
\end{equation}
where 
\begin{equation}
\mathrm{dilog}\left( z\right) =\mathrm{Li}_2\left( 1-z\right)
=\sum\limits_{k=1}^\infty \frac{\left( 1-z\right) ^k}{k^2}  \label{dil}
\end{equation}
is a dilogarithm function \cite{JST}.

At zero temperature $T=0$ the distribution function (\ref{f}) of a Fermi gas
is replaced by the Heaviside step function 
\begin{equation}
f_p=\Theta \left( p-p_F\right) =\Theta \left( \varepsilon -\varepsilon
_F\right)  \label{f1}
\end{equation}
where $p_F$ is the Fermi momentum and $\mu |_{T=0}=\varepsilon _F=ap_F^q$ is
the Fermi energy. Then, according to (\ref{na}) and , we find the particle
number density 
\begin{equation}
n=\Sigma _q\frac{\varepsilon _F}a=\Sigma _q\,p_F^q  \label{z1}
\end{equation}
and the energy density 
\begin{equation}
E=P=\Sigma _q\frac{\varepsilon _F^2}{2a}=\frac{\Sigma _q}2ap_F^{2q}
\label{en1}
\end{equation}
that implies 
\begin{equation}
P=E=\frac 1{2\Sigma _q}an^2  \label{p1q}
\end{equation}
Particularly, in 3 dimensions it is 
\begin{equation}
P=E=\frac{3\pi ^2}\gamma an^2  \label{p13}
\end{equation}
while in 2 dimensions it is\textrm{\ } 
\begin{equation}
P=E=\frac{2\pi }\gamma an^2  \label{p12}
\end{equation}
The 'absolute stiff'' matter at finite temperature and in 3D space is a
substance considered in astrophysical problems \cite{RR74,KB96,O,BE,BFM}.
Let us consider it in more detail.

\section{Fermionic absolute stiff matter}

Let us consider an 'absolute stiff' matter (\ref{sti}) in 3 dimensional
space. As we already have seen, this matter can be modeled by an ideal gas
of free quasi-particles with the energy spectrum (\ref{e11}). It can be
either a Fermi gas or a Bose gas, and its thermodynamical functions are
proportional to thermodynamical functions of nonrelativistic Fermi or Bose
gas in two dimensions because the latter has the same 'absolute stiff' EOS\ $%
P=E$ \cite{A92,T94}. However, the energy spectrum of 'absolute stiff'
quasi-particles (\ref{e11}) differs from the energy spectrum of ordinary
nonrelativistic particles (\ref{e112}), and we need to obtain all formulas
in detail and find exact coefficients.

Choosing the sign ''$+$'' in (\ref{f}) for fermions, defining 
\begin{equation}
x=\frac{m^4}{p^3T}  \label{def}
\end{equation}
and substituting (\ref{e11}) in (\ref{n}) and (\ref{p0}), we find the
particle number density 
\begin{equation}
n=\frac{\gamma m^2T}{6\pi ^2}\int\limits_0^\infty \frac{dx}{\exp \left(
x-\mu /T\right) +1}  \label{nf0}
\end{equation}
and the pressure and energy density 
\begin{equation}
E=P=\frac{\gamma m^2T^2}{6\pi ^2}\int\limits_0^\infty \frac{xdx}{\exp \left(
x-\mu /T\right) +1}  \label{pf0}
\end{equation}

All integrals are calculated in the explicit form or through special
functions, and we immediately find the particle number density 
\begin{equation}
n=\frac{\gamma m^2}{6\pi ^2}T\ln \left[ 1+\exp \left( \frac \mu T\right)
\right]  \label{nf1}
\end{equation}
and the pressure 
\begin{equation}
P=\frac{\gamma m^2}{6\pi ^2}\left\{ \frac 12\mu ^2+\frac{\pi ^2}6T^2+T^2%
\mathrm{dilog}\left[ 1+\exp \left( \frac{-\mu }T\right) \right] \right\}
\label{pf1}
\end{equation}
Formula (\ref{nf1}) also implies 
\begin{equation}
\mu =T\ln \left[ \exp \left( \frac{6\pi ^2n}{\gamma m^2T}\right) -1\right]
=T\ln \left[ \exp \left( \frac{\varepsilon _F}T\right) -1\right]  \label{mf1}
\end{equation}
where 
\begin{equation}
\varepsilon _F=\frac{6\pi ^2n}{\gamma m^2}  \label{ef1}
\end{equation}
is the Fermi energy.

At low temperature ($T\ll \varepsilon _F$) the chemical potential (\ref{mf1}%
) tends to a constant limit 
\begin{equation}
\mu \simeq \varepsilon _F  \label{mf2}
\end{equation}
and the pressure (\ref{pf1}) is approximated as 
\begin{equation}
P\simeq \frac{\gamma m^2}{6\pi ^2}\left( \frac 12\varepsilon _F^2+\frac{\pi
^2}6T^2\right)   \label{pf2}
\end{equation}
giving 
\begin{equation}
P=\frac{\gamma m^2}{12\pi ^2}\varepsilon _F^2=\frac \gamma {12\pi ^2}\frac{%
p_F^6}{m^2}  \label{pf22}
\end{equation}
at zero temperature, where 
\begin{equation}
p_F=\left( \frac{6\pi ^2n}\gamma \right) ^{1/3}  \label{pff}
\end{equation}
is the Fermi momentum. Hence, we find an important relation 
\begin{equation}
P=\frac{3\pi ^2}{\gamma m^2}n^2  \label{pf99}
\end{equation}
that coincides with (\ref{p13}) if $a=1/m^2$.

At high temperature $T\gg \varepsilon _F\,$ formula (\ref{mf1}) yields 
\begin{equation}
\mu \rightarrow T\ln \left( \frac{\varepsilon _F}T\right) \ll -T  \label{mf3}
\end{equation}
implying that the chemical potential is definitely negative. Hence, formula (%
\ref{nf1}) also implies 
\begin{equation}
n\simeq \frac{\gamma m^2}{6\pi ^2}T\exp \left( \frac \mu T\right)
\label{nf3}
\end{equation}
while the pressure (\ref{pf1}) tends to asymptotic 
\begin{equation}
P\rightarrow \frac{\gamma m^2}{6\pi ^2}T^2\exp \left( \frac \mu T\right)
\label{pf3}
\end{equation}
that corresponds to the standard EOS of classical Maxwell-Boltzmann gas 
\begin{equation}
P=nT  \label{pf4}
\end{equation}
whose particle number density 
\begin{equation}
n=\frac{\gamma m^2}{6\pi ^2}T\exp \left( \frac \mu T\right)  \label{nm}
\end{equation}
and pressure 
\begin{equation}
E=P=\frac{\gamma m^2T^2}{6\pi ^2}T^2\exp \left( \frac \mu T\right)
\label{pm}
\end{equation}
are obtained by formulas (\ref{n}) and (\ref{p0}) with the distribution
function 
\begin{equation}
f_p=\exp \left( \frac{\mu -\varepsilon _p}T\right)  \label{fm}
\end{equation}

If we consider 'absolute stiff' thermal excitations whose number is not
conserved, it is necessary to put the chemical potential equal to zero $\mu
=0$. According to (\ref{nf1}) and (\ref{pf1}), their particle number density
and pressure are 
\begin{equation}
n_{exc}=\frac{\gamma \ln 2}{6\pi ^2}m^2T  \label{nft}
\end{equation}
and 
\begin{equation}
P^{exc}=\frac{\gamma m^2}{72}T^2  \label{pft}
\end{equation}
Now 
\begin{equation}
P^{exc}=\frac{\pi ^4}{2\ln ^22}\frac{n_{exc}^2}{\gamma m^2}  \label{pft2}
\end{equation}
that differs from the relevant identity for the cold Fermi gas (\ref{pf99}),
however, proportionality $P\sim n^2/m^2$ is the same in both cases. 

\section{Bosonic absolute stiff matter}

Choosing the sign ''$-$'' in (\ref{f}) we obtain the thermodynamical
functions of bosonic 'absolute stiff' matter 
\begin{equation}
n=\frac{\gamma m^2T}{6\pi ^2}\int\limits_0^\infty \frac{dx}{\exp \left(
x-\mu /T\right) -1}  \label{nb0}
\end{equation}
\begin{equation}
E=P=\frac{\gamma m^2T^2}{6\pi ^2}\int\limits_0^\infty \frac{xdx}{\exp \left(
x-\mu /T\right) -1}  \label{pb0}
\end{equation}
All integrals a obtained in the explicit form or through special functions,
and we immediately find the particle number density 
\begin{equation}
n=-\frac{\gamma m^2}{6\pi ^2}T\ln \left( 1-\exp \frac \mu T\right)
\label{nb1}
\end{equation}
and the pressure 
\begin{equation}
P=\frac{\gamma m^2}{6\pi ^2}\left\{ \frac 12\mu ^2+\frac{\pi ^2}6T^2+T^2%
\mathrm{dilog}\left[ \exp \left( \frac{-\mu }T\right) \right] -\mu T\ln
\left[ 1-\exp \left( \frac \mu T\right) \right] \right\}  \label{pb1}
\end{equation}
where function $\mathrm{dilog}\left( x\right) $ is defined by (\ref{dil}).
Formula (\ref{nb1}) allows to determine the chemical potential of the Bose
gas 
\begin{equation}
\mu =T\ln \left[ 1-\exp \left( -\frac{6\pi ^2n}{\gamma m^2T}\right) \right]
=T\ln \left[ 1-\exp \left( -\frac{T_c}T\right) \right]  \label{mb1}
\end{equation}
where the characteristic temperature is 
\begin{equation}
T_c=\frac{6\pi ^2n}{\gamma m^2}  \label{tc}
\end{equation}
The chemical potential $\left| \mu \right| $ is growing with the growth of
temperature, it is always negative and attains zero $\mu \rightarrow 0$ only
in the limit $T\rightarrow 0$. So, there is no Bose-Einstein condensation of
the 'absolute stiff' bosons.

At low temperature $T\ll T_c$ the chemical potential (\ref{mb1}) tends to
zero 
\begin{equation}
\mu \simeq -T\exp \left( -\frac{T_c}T\right) \rightarrow 0  \label{mb2}
\end{equation}
while the pressure (\ref{pb1}) is approximated by formula 
\begin{equation}
P\cong \frac{\gamma m^2}{36}T^2  \label{pb2}
\end{equation}
It should be noted that the number of particles (\ref{nb1}) is divergent 
\begin{equation}
n=\frac{\gamma m^2}{6\pi ^2}T\ln \left( \frac T{-\mu }\right)  \label{nb2}
\end{equation}
and there is no Bose-Einstein condensation because the chemical potential (%
\ref{mb2}) never attains zero level.

At high temperature $T\gg T_c$ the chemical potential (\ref{mb1}) is
approximated so 
\begin{equation}
\mu \rightarrow -T\ln \left( \frac T{T_c}\right) \ll -T  \label{mb3}
\end{equation}
while formula (\ref{nb1}) also implies 
\begin{equation}
n\simeq \frac{\gamma m^2}{6\pi ^2}T\exp \left( \frac \mu T\right)
\label{nb3}
\end{equation}
and the pressure (\ref{pb1}) tends to 
\begin{equation}
P\rightarrow \frac{\gamma m^2}{6\pi ^2}T^2\exp \left( \frac \mu T\right)
\label{pb3}
\end{equation}
that again gives the EOS of Maxwell-Boltzmann gas $P=nT$.

\section{Anyonic matter and heat capacity}

For the general anyon distribution function 
\begin{equation}
f_p=\frac 1{\exp \left[ (\varepsilon _p-\mu )/T\right] +\sigma }  \label{fa}
\end{equation}
with arbitrary $\sigma $ and 'absolute stiff' energy spectrum (\ref{e11})
formulas (\ref{n}) and (\ref{p0}) yield the partcile number density

\begin{equation}
n=\frac{\gamma m^2}{6\pi ^2}\frac T\sigma \ln \left[ 1+\sigma \exp \left(
\frac \mu T\right) \right]  \label{nfa}
\end{equation}
and pressure 
\begin{equation}
E=P=\frac{\gamma m^2}{6\pi ^2\sigma }\left\{ \frac{\mu ^2}2+\left( \frac{\ln
^2\sigma }2+\frac{\pi ^2}6\right) T^2+T^2\mathrm{dilog}\left[ 1+\frac
1\sigma \exp \left( \frac{-\mu }T\right) \right] +\mu T\ln \sigma \right\}
\label{pfa}
\end{equation}
that at $\sigma =\pm 1$ are reduced to (\ref{nf1})-(\ref{pf1}) and (\ref{nb1}%
)-(\ref{pb1}), respectively.

The the entropy density $S$ and the heat capacity $C_V$ are defined by
formulas 
\begin{equation}
S=-\frac{\partial \left( T\ln Z\right) }{\partial T}=\frac{\partial P}{%
\partial T}\qquad C_V=T\frac{\partial S}{\partial T}  \label{ent}
\end{equation}
Substituting (\ref{pfa}) in (\ref{ent}), we get 
\begin{equation}
S=\frac{\gamma m^2}{6\pi ^2\sigma }\left\{ T\left( \ln ^2\sigma +\frac{\pi ^2%
}3\right) +2T\mathrm{dilog}\left[ 1+\frac 1\sigma \exp \left( \frac{-\mu }%
T\right) \right] -\mu \ln \left[ \frac 1\sigma +\frac 1{\sigma ^2}\exp
\left( \frac{-\mu }T\right) \right] \right\}  \label{sa}
\end{equation}
and 
\begin{eqnarray}
C_V &=&\frac{\gamma m^2}{6\pi ^2\sigma }\left\{ T\left( \ln ^2\sigma +\frac{%
\pi ^2}3\right) +2T\mathrm{dilog}\left[ 1+\frac 1\sigma \exp \left( \frac{%
-\mu }T\right) \right] -\right.  \nonumber \\
&&\ \ \ \left. \qquad -\frac{\mu ^2/T}{1+\sigma \exp \left( \mu /T\right) }%
-2\mu \ln \left[ 1+\frac 1\sigma \exp \left( \frac{-\mu }T\right) \right]
\right\}  \label{ca}
\end{eqnarray}

Particlularly, at $\sigma =1$ we have the entropy density of 'absolute
stiff' Fermi gas 
\begin{equation}
S_F=\frac{\gamma m^2}{6\pi ^2}\left\{ \frac{\pi ^2}3T+2T\mathrm{dilog}\left[
1+\exp \left( -\frac \mu T\right) \right] -\mu \ln \left[ 1+\exp \left(
-\frac \mu T\right) \right] \right\}  \label{fe}
\end{equation}
and at $\sigma =-1$ we have the entropy density of 'absolute stiff' Bose gas 
\begin{equation}
S_B=\frac{\gamma m^2}{6\pi ^2}\left\{ \frac{2\pi ^2}3T+2T\mathrm{dilog}%
\left[ 1-\exp \left( -\frac \mu T\right) \right] +\mu \ln \left[ \exp \left( 
\frac{-\mu }T\right) -1\right] \right\}  \label{bo}
\end{equation}
Substituting (\ref{fe}) in (\ref{ent}) we find the heat capacity of
'absolute stiff' Fermi gas 
\begin{equation}
C_F=\frac{\gamma m^2}{6\pi ^2}\left\{ \frac{\pi ^2}3T+2T\mathrm{dilog}\left[
1+\exp \left( \frac{-\mu }T\right) \right] -2\mu \ln \left[ 1+\exp \left( 
\frac{-\mu }T\right) \right] -\frac{\mu ^2/T}{1+\exp \left( \mu /T\right) }%
\right\}  \label{hf}
\end{equation}
Substituting (\ref{bo}) in (\ref{ent}), we find the heat capacity of the
'absolute stiff' Bose gas 
\begin{equation}
C_B=\frac{\gamma m^2}{6\pi ^2}\left\{ \frac{2\pi ^2}3T-2T\mathrm{dilog}%
\left[ 1-\exp \left( -\frac \mu T\right) \right] +2\mu \ln \left[ 1-\exp
\left( \frac{-\mu }T\right) \right] +\frac{\mu ^2/T}{1-\exp \left( \mu
/T\right) }\right\}  \label{hb}
\end{equation}
At low temperature ($T\ll \varepsilon _F$ and $T\ll T_c$) the heat capacity
of both fermionic and bosonic 'absolute stiff' matter behaves as 
\begin{equation}
C_F\rightarrow C_B\rightarrow \frac{\gamma m^2}{18}T  \label{hfb}
\end{equation}
At high temperature ($T\gg \varepsilon _F$ and $T\gg T_c$) the heat capacity
becomes exponentially small 
\begin{equation}
C_F\rightarrow C_B\rightarrow \frac{\gamma m^2}{6\pi ^2}\frac{\mu ^2}T\exp
\left( \frac \mu T\right)  \label{hbb}
\end{equation}

According to formulas (\ref{pft}) and (\ref{ent}), the heat capacity of
fermionic 'absolute stiff' thermal excitations (that have zero chamical
potential $\mu =0$) is always 
\begin{equation}
C_F^{exc}=\frac{\gamma m^2}{36}T  \label{hfb2}
\end{equation}

\section{Conclusion}

The particle number density $n$ and the pressure $P$ of 'absolute stiff'
matter with the equation of state $P=E$ (\ref{sti}) at finite temperature
are given by formulas (\ref{pf1})-(\ref{mf1}) and (\ref{pb1})-(\ref{mb1})
for fermions and bosons, respectively. There is a critical temperature%
\textrm{\ } 
\begin{equation}
T_c=\frac{6\pi ^2n}{\gamma m^2}  \label{temp}
\end{equation}
that characterizes the behavior of Fermi and Bose gases. At low temperature (%
$T\ll T_c$) the fermionic pressure is approximated by formula (\ref{pf2}),
while the bosonic pressure is (\ref{pb2}). A mixture of fermions and bosons
at low temperature will have the pressure 
\begin{equation}
P\cong \frac{3\pi ^2}{\gamma _Fm_F^2}n^2+\frac{\gamma _Fm_F^2+\gamma _Bm_B^2%
}{36\pi ^2}T^2  \label{pfb}
\end{equation}
At high temperature ($T\gg T_c$) both fermions and bosons behave as ideal
Maxwell-Boltzmann gases with $P=nT$, although the constraint (\ref{sti})
remains valid. The heat capacity of 'absolute stiff' Fermi and Bose gases is
given by formulas (\ref{hf})-(\ref{hb}) and at low temperature it is
approximated by the same linear asymptotic (\ref{hfb}).

The EOS of 'absolute stiff' Fermi gas at zero temperature is characterized
by proportionality (\ref{pf99}): 
\begin{equation}
P\sim n^2  \label{psti}
\end{equation}
Although this formula is applied only to a system of fermions at zero
temperature, it has can be used in various problems concerned with neutron
stars. The central density of neutron $n_c$ exceeds several times the normal
nuclear density $n_{nm}$, while the regular nuclear EOS can be accurately
estimated when the density is not so high. Some researches appeal to the
model of 'absolute stiff' matter when $n$ is larger some fiducial density $%
n_{\perp }$ around $4n_{nm}$ \cite{KB96}. So, we can calculate the pressure $%
P_{\perp }$ in the frames of regular nuclear EOS at zero temperature. Then,
substituting $n_c$ and $n_{\perp }$ in (\ref{psti}) we immediately determine
the pressure in the center of the star 
\begin{equation}
P_c=P_{\perp }\frac{n_c^2}{n_{\perp }^2}  \label{fed}
\end{equation}
At finite temperature this constraint should be replaced by 
\begin{equation}
P_c=P_{\perp }\frac{n_c^2+\alpha T_c^2}{n_{\perp }^2+\alpha T_{\perp }^2}%
\qquad \alpha =\frac{\gamma _Fm_F^2\left( \gamma _Fm_F^2+\gamma
_Bm_B^2\right) }{108\pi ^4}  \label{fed2}
\end{equation}
that also includes the central temperature $T_c$ and the temperature $%
T_{\perp }$ at the boundary between the envelope and the core with 'absolute
stiff' EOS.

Thus, the EOS $P=wE$ (\ref{xi}) with arbitrary constant $w$ can be modeled
by a system of free quasi-particles that have energy spectrum (\ref{e0}),
without regard which statistics they obey. Particularly, we can consider
fermionic or bosonic stars containing an ideal gas of quasi-particles that
constitute exotic matter with $P<-E$ as an alternative to the Chaplygin gas 
\cite{Chap}. It is the subject for further research.

\newpage


\end{document}